\documentclass[12pt,preprint]{aastex}

\shorttitle{Donor Star of SS~433}
\shortauthors{Gies et al.}

\begin{document}

\received{}
\accepted{}
 
\title{The Spectrum of the Mass Donor Star in SS~433}

\author{D. R. Gies\altaffilmark{1},
        W. Huang\altaffilmark{1}, and 
        M. V. McSwain}

\affil{Center for High Angular Resolution Astronomy, 
Department of Physics and Astronomy \\
Georgia State University, Atlanta, GA  30303\\
Electronic mail: gies@chara.gsu.edu, huang@chara.gsu.edu,
mcswain@chara.gsu.edu}

\altaffiltext{1}{Visiting Astronomer, University of Texas McDonald Observatory.}

\slugcomment{Submitted to ApJL.}
\paperid{}


\begin{abstract}
We present results from a short series of blue, moderate 
resolution spectra of the microquasar binary, SS~433. 
The observations were made at a time optimized to find the 
spectrum of the donor star, i.e., when the donor was in the 
foreground and well above the plane of the obscuring disk. 
In addition to the well-known stationary and jet emission 
lines, we find evidence of a weak absorption spectrum that 
resembles that of an A-type evolved star. 
These lines display radial velocity shifts opposite to 
those associated with the disk surrounding the compact star, 
and they appear strongest when the disk is maximally eclipsed. 
All these properties suggest that these absorption lines 
form in the atmosphere of the hitherto unseen mass donor star 
in SS~433.   The radial velocity
shifts observed are consistent with a mass ratio
$M_X / M_O = 0.57 \pm 0.11$ and masses of $M_O = (19\pm 7)~M_\odot$ and
$M_X = (11 \pm 5)~M_\odot$.  These results indicate that the system 
consists of an evolved, massive donor and a black hole 
mass gainer. 
\end{abstract}

\keywords{stars: early-type --- 
 stars: individual (SS~433; V1343~Aql) --- X-rays: binaries}


\section{Introduction}                              

SS~433 is still one of the most mysterious of the X-ray binaries 
even after some 25 years of observation \citep{mar84,zwi89,gie02}. 
We know that the mass donor feeds an enlarged accretion disk surrounding a
neutron star or black hole companion, and a small portion of this inflow is 
ejected into relativistic jets that are observed in optical and X-ray
emission lines and in high resolution radio maps.  There are two basic
timescales that control the spectral appearance and system dynamics,
a 162~d disk and jet precessional cycle and a 13~d orbital period.
The mass function derived from the \ion{He}{2} $\lambda 4686$ emission line 
indicates that the donor star mass is in excess of $8~M_\odot$ \citep{fab90}, 
and the donor is probably a Roche-filling, evolved star \citep{kin00}.
However, the spectral signature of this star has eluded detection.
This is probably because the binary is embedded in an
expanding thick disk that is fed by the wind from the super-Eddington 
accretion disk \citep{zwi91}.  The outer regions of this equatorial thick disk 
have been detected in high resolution radio measurements by 
\citet{par99} and \citet{blu01}.  We recently showed 
how many of the properties of the ``stationary'' emission lines 
can be explained in terms of a disk wind \citep{gie02}.  

The task of finding the spectrum of the donor is
crucial because without a measurement of its orbital motion,
the mass of the relativistic star is unknown.   The best opportunity to
observe the flux from the donor occurs at the precessional
phase when the disk normal is closest to our line of sight and 
the donor star appears well above the disk plane near the 
donor inferior conjunction orbital phase \citep{gie02}.   
This configuration occurs only a few nights each year 
for ground-based observers.   The choice of spectral region 
is also important.  \citet{gor98b} found that the regular eclipse and
precessional variations seen clearly in the blue are lost in the
red due to an erratically variable flux component. 
Thus, we need to search for the donor spectrum blueward of
the $R$-band in order to avoid this variable component.   
On the other hand, the color variations observed during eclipses 
suggest that the donor is cooler than
the central portions of the disk \citep{ant87,gor97}.
Thus, the disk will tend to contribute a greater fraction of the
total flux at lower wavelengths.   The best compromise is in the
blue where there are a number of strong
absorption lines in B- and later-type stars.

Here we present the results of a blue spectral search for 
the donor's spectrum made during an optimal disk and
orbital configuration in 2002 June (\S2).  We first discuss the 
dominant emission features formed in the jets and disk wind (\S3). 
We then focus on a much weaker set of absorption lines 
(\S4), and we present arguments linking these to the photosphere 
of the donor star.  


\section{Observations}                              

The blue spectra of SS~433 were obtained on three consecutive 
nights, 2002 June 5 -- 7, with the Large Cassegrain Spectrograph  
(LCS) on the 2.7-m Harlan J.\ Smith Telescope at the 
University of Texas McDonald Observatory \citep{coc02}.  
These dates correspond to precessional phases between 
$\Psi = 0.998$ and 0.011 (where $\Psi = 0.0$ corresponds to the time 
when the jets are closest to our line of sight and their 
emission lines attain their extremum radial velocities) 
and to orbital phases between $\phi =0.012$ and 0.178 (where 
mid-eclipse and donor inferior conjunction occur at
$\phi = 0.0$) according to the phase relations adopted in 
\citet{gie02}.  The spectra were made in first order 
with the \#46 grating (1200 grooves mm$^{-1}$, blazed at 4000 \AA ), 
and they cover the wavelength range between 4060 and 4750 \AA .
We used a long slit configuration with a slit width of $2\farcs0$,
which corresponds to a projected width of 2 pixels FWHM on the detector,
a $800\times 800$ format TI CCD with 15$\mu$m square pixels. 
The reciprocal dispersion is 0.889 \AA ~pixel$^{-1}$
and the spectral resolving power is $\lambda / \triangle\lambda = 2500$. 
Unfortunately, the weather was partially cloudy throughout 
the run, and we were only able to obtain a few consecutive, 20 minute 
exposures on each night.   We also obtained a full suite of calibration 
bias, dome flat field, and argon comparison frames throughout the run.

The spectra were extracted and calibrated
using standard routines in IRAF\footnote{IRAF is distributed by the
National Optical Astronomy Observatories, which is operated by
the Association of Universities for Research in Astronomy, Inc.,
under cooperative agreement with the National Science Foundation.}.
We co-added all the consecutive spectra from 
a given night to increase the signal-to-noise ratio (S/N). 
The co-added spectra were then rectified to a unit continuum 
by the fitting of line-free regions.
Note that this rectification process arbitrarily removes 
the continuum flux variations that occurred in SS~433 during the eclipse.
All the spectra were then transformed to a common heliocentric 
wavelength grid for ease of comparison.  The final S/N ratios in the continuum 
at the long-wavelength end of the spectra are 31, 16, and 40 pixel$^{-1}$
for the three consecutive nights, respectively. 


\section{Emission Line Spectrum}                    

Our three spectra are shown in chronological sequence in Figure~1. 
This part of the spectrum is dominated by the familiar strong 
emission lines of H$\delta$, H$\gamma$, and \ion{He}{2} $\lambda 4686$ 
\citep{mur80,panfab97,fab97b}.  Most of these lines appear stronger
in the first, mid-eclipse spectrum made when the continuum 
flux was low.   We begin by examining the radial velocity and 
intensity variations of the emission lines before turning 
to the weaker absorption features in the next section. 

\placefigure{fig1}     

The only strong jet emission feature in this spectral range 
at this time is the blueshifted H$\beta -$ line (formed in the 
approaching jet), which is found just longward of H$\gamma$.  
This spectral range should also contain redshifted jet components from 
the upper Balmer sequence (from the Balmer limit at $\approx 4267$ \AA 
~to H$\epsilon$ at $\approx 4645$ \AA ), but, if present, 
they are too weak to detect in our spectra.   We measured 
the radial velocity of the H$\beta -$ emission feature by fitting 
a single Gaussian in each case.  These measurements are partially 
compromised by the appearance of multiple sub-peaks in some cases
(see the final profile that contains a redshifted, sub-peak 
corresponding in radial velocity to the ``bullet'' that 
appeared on the previous night) and by the possible existence 
of very weak \ion{He}{1} $\lambda 4387$ emission.  
Nevertheless, these measurements show the 
familiar oscillations in the jet velocities due to the 
tidal ``nodding'' of the accretion disk \citep{ver93a,gie02}. 
Table~1 lists the heliocentric Julian date for the 
mid-time of each observation, the observed 
H$\beta -$ radial velocity as $z = V_r / c$ 
(estimated errors $\pm 0.001$), and the predicted 
$z$ value based upon the extrapolation of the fit to 
the H$\alpha$ jet velocities observed between 1998 and 1999 
\citep{gie02}.   The good agreement between the 
observed and predicted motions suggests that our
jet precessional velocity fit is still reliable 
3 years after the last H$\alpha$ observations. 

\placetable{tab1}      

The stationary emission lines generally fall into two 
categories: lines like \ion{He}{2} $\lambda 4686$ that
form in a region symmetric about the center of the accretion disk 
and that show orbital radial velocity curves of the form 
$-K_1 \sin (2\pi \phi)+V_1$, and features like the 
hydrogen Balmer lines that probably form in a large 
volume in the disk's wind and have a radial velocity 
variation of the form $K_2 \cos (2\pi \phi)+V_2$ \citep{gie02}.
We measured the radial velocities of the stationary lines 
in our spectra to confirm their radial velocity behavior. 
The profiles were obtained by Gaussian fitting in all cases 
except for the \ion{N}{3} + \ion{C}{3} complex near 4644 \AA ,
where we measured relative shifts by cross-correlating the 
profiles for the second and third nights against that from the 
first night.  Our results are listed in Table~2
(errors are $\pm 30$ km~s$^{-1}$). 
All the lines showed the blueward motion expected in this 
phase interval, but with a larger amplitude than predicted. 
The largest decline was found in \ion{N}{3} + \ion{C}{3} 
and \ion{He}{2} $\lambda 4686$, which decreased 
in radial velocity by $\approx 214$ km~s$^{-1}$ over the duration 
of observations.   \citet{fab90} estimate that 
the disk semiamplitude is $K_1 \approx 
175$ km~s$^{-1}$, much smaller than we observed. 
The same situation occurred in lines of the other group, 
H$\delta$, H$\gamma$, and \ion{He}{1} $\lambda 4471$ which 
declined by some 108 km~s$^{-1}$ (compared to the 
expected $K_2 \approx 64$ km~s$^{-1}$; \citet{gie02}). 
We speculate that this difference is due to the 
emergence of the approaching portion of the disk after the eclipse,
which biases the radial velocities to more negative values. 

\placetable{tab2}      

The other clue about the origin of the stationary lines comes
from their intensity variations during the eclipse.  If the 
emission source has a constant flux, then during eclipse
the line will appear proportionally stronger relative to the 
lower continuum flux.  However, if the emission source is 
also (partially) eclipsed, then the relative strength in 
rectified intensity remains approximately constant in eclipse. 
We show in Table~3 the rectified intensities 
of the emission lines for the first two nights 
relative to that on the final night. 
The final column gives the predicted constant emission flux variation 
relative to the last night based upon the 
$V$-band light curve (\citet{gor98a}; see their Fig.~7b)
and the $B-V$ color variations (\citet{gor97}; see their Fig.~3). 
We find that most of the line intensities
decreased by a factor of $2.4\times$
between the first (mid-eclipse) and last (out-of-eclipse) observations,
compared to an expected decrease of $1.9\times$ for a constant 
emission flux source.  This difference is probably not significant 
since the intrinsic system flux does vary on short timescales, 
but the result does suggest that most of the lines form in regions large 
compared to the continuum-forming part of the accretion disk 
(so that the continuum-forming inner disk is eclipsed while 
the much larger line forming region suffers only minor occultation).  
The main exception is the \ion{N}{3} + \ion{C}{3} line which 
varies little in rectified intensity and must therefore form 
co-spatially with the continuum in the inner, hot part 
of the accretion disk.  The other possible exception is the 
\ion{He}{2} $\lambda 4686$ line.  \citet{gor97} and \citet{fab97a}
observed profile shape variations during the eclipse that 
we also find (for example, from single- to double-peaked
with egress from the eclipse), and they argue that the 
feature has two components, one formed in gas close to the disk 
center (eclipsed) and one formed in a larger volume 
(not eclipsed).   This assessment agrees with the 
fact that, after the \ion{N}{3} + \ion{C}{3} line, the 
\ion{He}{2} $\lambda 4686$ feature shows the second smallest 
decrease in rectified flux, suggesting that its line forming 
region is more occulted during the eclipse than is the 
case for the H and \ion{He}{1} lines (formed in the larger
disk wind). 

\placetable{tab3}      


\section{Absorption Line Spectrum}                  

We were struck in comparing the individual spectra 
by the similarity of patterns of what might first be considered 
``noise'' in the continuum between the strong emission 
lines.  We formed a global average spectrum with each 
spectrum weighted by the square of its S/N ratio, and 
an expanded version of this average spectrum is shown 
in Figures 2 and 3.  We show below the average SS~433 plot
examples of stellar spectra that could correspond to the
spectral type of the donor star.  The two B-star spectra 
were obtained from the atlas of \citet{wal90}, and 
we obtained the spectrum of the A-supergiant, HD~148743, 
with the LCS.  All these spectra
were Gaussian smoothed to an effective 2 pixel resolution 
of 1.78 \AA ~FWHM in order to compare them at the 
same instrumental resolution.   We see that there is 
a significant (and largely unresolved) system of 
absorption lines in SS~433 that has eluded detection  
in earlier work.   The line patterns bear little resemblance 
to those in B-supergiants but there are a number of 
features in common with the A-supergiant
($T_{\rm eff} = 7800$~K; \citet{ven95}).  
We show some preliminary identifications 
of the absorption lines based upon lists in \citet{bal80}
and \citet{ven95}. 

\placefigure{fig2}     

\placefigure{fig3}     

We measured the radial velocity of the absorption system by 
cross-correlating sub-sections of the spectrum containing 
the deepest features with the same in the first spectrum of the set.
These relative velocities are listed in the final column of 
Table~2.  (We found that the absorption lines in the first spectrum had 
a cross-correlation velocity of $-39\pm 20$ km~s$^{-1}$ compared to the
rest frame spectrum of HD~148743, so 
adding this value to the velocities in Table~2 gives an
estimate of the absolute velocities.) 
Unlike the emission lines, the absorption 
lines moved significantly redward during the run, 
as expected for the donor star. 
 
Direct inspection of Figure~1 suggests that the absorption line
spectrum grew fainter with egress from the eclipse, as predicted 
for the donor's spectrum.  It is difficult to measure the change in 
individual lines, so we measured the strength of the 
cross-correlation of the absorption lines
with those in HD~148743.  The relative 
cross-correlation intensities are listed in the second last 
column of Table~3 (errors of $\pm 0.5$).   The 
absorption depths weakened in the same way as
the emission intensities, suggesting that both became diluted 
by the emerging flux of the disk.   

Both the radial velocity and intensity variations indicate that
the absorption lines form in the long-sought donor star.  
If they were formed in the interstellar medium (the case 
of the absorptions at 4428, 4501, and 4726 \AA ), then we would observe 
no velocity or intensity variations through these eclipse phases. 
Similar arguments rule out formation in an extended 
``shell'' surrounding the binary system.  

The absorption line velocities are insufficient for a 
general orbital solution, but we can make a restricted 
solution if we assume that the orbit is circular and 
we adopt the orbital period and epoch of mid-eclipse from 
\citet{gor98a}.  Then we can solve for two parameters,
the systemic velocity, $V_0$, and orbital semiamplitude, $K_O$,
from our 3 radial velocity measurements of the optical star.  
We used the orbital fitting code of \citet{mor74} to find
$V_0= (-44 \pm 9)$ km~s$^{-1}$ and $K_O = (100 \pm15)$ km~s$^{-1}$ with residual 
errors of 9 km~s$^{-1}$.  \citet{fab90} find the semi-amplitude 
of the disk is $K_X = (175\pm 20)$ km~s$^{-1}$
based upon the \ion{He}{2} $\lambda 4686$ radial velocity curve.
They adopt the system inclination from the ``kinematical'' model of the jets 
to arrive at a mass relation, $M_O/(1+q)^2 = (7.7\pm 2.7)~M_\odot$ 
where the mass ratio is $q=M_X/M_O$.  We can now combine
the semiamplitudes to estimate the mass ratio, 
$q=K_O/K_X = 0.57\pm 0.11$, and the resulting masses
are $M_O=(19\pm 7)~M_\odot$ and $M_X=(11\pm 5)~M_\odot$. 
Thus, our results suggest that the companion is a black hole. 

The absorption line depths strengthen from main sequence to
supergiant in the A stars, and they appear sufficiently strong 
in the first eclipse spectrum to rule out a main sequence class. 
The donor star probably fills the critical Roche surface, 
and our estimate of the mass ratio indicates a Roche volume 
radius of $(31\pm 3)~R_\odot$, consistent with a supergiant class. 
Thus, our results for SS~433 support the evolutionary scenario
described by \citet{kin00} in which mass transfer is occurring on 
a thermal timescale as the donor crosses the Hertzsprung gap. 

We can estimate the magnitude difference
between the star and disk, $\triangle B$, 
based upon the apparent line depths.  Suppose that during the central 
eclipse a disk flux of $F_1$ is occulted while a disk flux of 
$F_2$ and donor star flux $F_\star$ remain visible. 
The stellar line depths will then 
appear diluted by a factor $F_\star/(F_2 + F_\star )$. 
The observed line depths during eclipse (relative to the 
spectrum of the A-supergiant) suggest that the dilution 
is minimal, so $F_2/ F_\star = 0$ to 1.  
Based upon the line intensity variations 
we observed (Table~3), the ratio of 
out-of-eclipse to mid-eclipse flux is $2.38 \pm 0.15$,
and thus, the disk to star flux ratio is 
$(F_1 + F_2) / F_\star = 1.4$ to 2.8 (or $\triangle B 
= 0.3$ to 1.4 mag).  A donor star this bright may appear to be 
in conflict with earlier results \citep{ant87,gie02}, 
but we suspect that the star is heavily obscured at 
other precessional and orbital phases so that the line 
spectrum is difficult to find (and thus the estimate 
of the donor star's flux based on the absence of the lines 
will to be too low). 

Clearly our results should be regarded as preliminary since 
we have only observed the absorption spectrum during this 
one eclipse event, and SS~433 is known to display spectroscopic 
variations on timescales unrelated to the orbit. 
Nevertheless, confirmation of our results (especially in 
spectra of higher S/N and resolution) would be of particular 
importance in determining the stellar parameters of the donor 
($T_{\rm eff}$, $\log g$, and abundance) and in 
refining the mass estimates.   We emphasize again that 
the successful detection of the donor spectrum is probably 
limited to times near $\Psi = 0$ and $\phi = 0$. 
The next opportunities will 
occur near 2003 April 28 and 2003 October 2. 


\acknowledgments

We are grateful to the staff of McDonald Observatory and 
especially Dr.\ Anita Cochran for their help. 
Support for this work was provided by NASA through grant number
GO-8308 from the Space Telescope Science Institute, which is
operated by the Association of Universities for Research in
Astronomy, Inc., under NASA contract NAS5-26555. 
Institutional support has been provided from the GSU College
of Arts and Sciences and from the Research Program Enhancement 
fund of the Board of Regents of the University System of Georgia,
administered through the GSU Office of the Vice President 
for Research.   We gratefully acknowledge all this support. 





\clearpage

\begin{figure}
\caption{The rectified spectra of SS~433 marked with labels for 
the prominent features. The spectra are shown chronologically 
from the first night ({\it upper}) to the last night ({\it lower}), 
and the spectrum for each night is offset by unity for clarity.
The orbital phases of observation, $\phi$, are listed above 
each spectrum.} 
\label{fig1}
\end{figure}

\begin{figure}
\caption{A detailed plot of the low wavelength portion of the 
co-added spectrum of SS~433 ({\it thick line}).  
The individual spectra were shifted to the rest frame 
prior to co-addition. 
Several examples of the predicted donor spectrum are 
shown below ({\it thin lines}): 
HD~148743 (A7~Ib), $\beta$~Ori (B8~Ia), and 
HD~148688 (B1~Ia).  All the spectral intensities were 
increased by a factor of 2 and offset in intensity 
for ease of comparison.  Preliminary identifications are given 
for a number of weak absorption lines in the spectrum of SS~433.
A plus sign following the identification indicates 
a blend of multiple lines.} 
\label{fig2}
\end{figure}

\begin{figure}
\caption{A detailed plot of the long wavelength portion of the 
co-added spectrum of SS~433 in the same format as Fig.~2.} 
\label{fig3}
\end{figure}


\clearpage


\begin{deluxetable}{lcc}
\tablewidth{0pc} 
\tablecaption{Jet Radial Velocity Measurements \label{tab1}}
\tablehead{
\colhead{Date} &
\colhead{Observed} &
\colhead{Predicted} \\
\colhead{(HJD - 2450000)} &
\colhead{$z$(H$\beta -$)} &
\colhead{$z$(H$\alpha -$)}
} 
\startdata
2430.7616\dotfill &  $-$0.099 &  $-$0.098 \\
2431.8226\dotfill &  $-$0.097 &  $-$0.095 \\
2432.9381\dotfill &  $-$0.103 &  $-$0.100 \\
\enddata
\end{deluxetable}

\clearpage


\begin{deluxetable}{lcccccc}
\tablewidth{0pc} 
\tablecaption{Stationary Line Radial Velocity Measurements \label{tab2}}
\tablehead{
\colhead{Date} &
\colhead{$V_r$(H$\delta$)} &
\colhead{$V_r$(H$\gamma$)} &
\colhead{$V_r$(\ion{He}{1})} &
\colhead{$V_r$(\ion{N}{3})\tablenotemark{a}} &
\colhead{$V_r$(\ion{He}{2})} &
\colhead{$V_r$(Abs.)\tablenotemark{a}} \\
\colhead{(HJD - 2450000)} &
\colhead{(km s$^{-1}$)} &
\colhead{(km s$^{-1}$)} &
\colhead{(km s$^{-1}$)} &
\colhead{(km s$^{-1}$)} &
\colhead{(km s$^{-1}$)} &
\colhead{(km s$^{-1}$)} 
} 
\startdata
2430.7616\dotfill &     169 &   204 & 264 & \phn\phn\phs0 &    \phs168 & 0 \\
2431.8226\dotfill &  \phn25 &   118 & 204 &     \phn$-$96 & \phn\phs50 & $38\pm 35$ \\
2432.9381\dotfill &  \phn77 &\phn80 & 157 &        $-$216 &  \phn$-$44 & $87\pm 15$ \\
\enddata
\tablenotetext{a}{Cross-correlation velocity relative to first observation.}
\end{deluxetable}

\clearpage


\begin{deluxetable}{lcccccccc}
\tablewidth{0pc} 
\tablecaption{Relative Line Intensities $F_l/F_c$ : $F_l/F_c (\phi = 0.178)$ \label{tab3}}
\tablehead{
\colhead{Orbital Phase} &
\colhead{H$\delta$} &
\colhead{H$\gamma$} &
\colhead{\ion{He}{1}} &
\colhead{\ion{N}{3}} &
\colhead{\ion{He}{2}} &
\colhead{H$\beta -$} &
\colhead{Abs.\tablenotemark{a}} &
\colhead{$B$ Light Curve\tablenotemark{b}}
} 
\startdata
0.012\dotfill & 2.28 & 2.53 & 2.47 & 1.20 & 2.16 & 2.44 & 2.6 & 1.91 \\
0.093\dotfill & 1.58 & 1.38 & 1.56 & 1.16 & 1.40 & 1.25 & 1.5 & 1.22 \\
\enddata
\tablenotetext{a}{Cross-correlation intensity relative to that at phase 0.178.}
\tablenotetext{b}{Prediction for constant $F_l$ relative to phase 0.178.}
\end{deluxetable}

\clearpage



\clearpage

\setcounter{figure}{0}
\begin{figure}[t]
\plotone{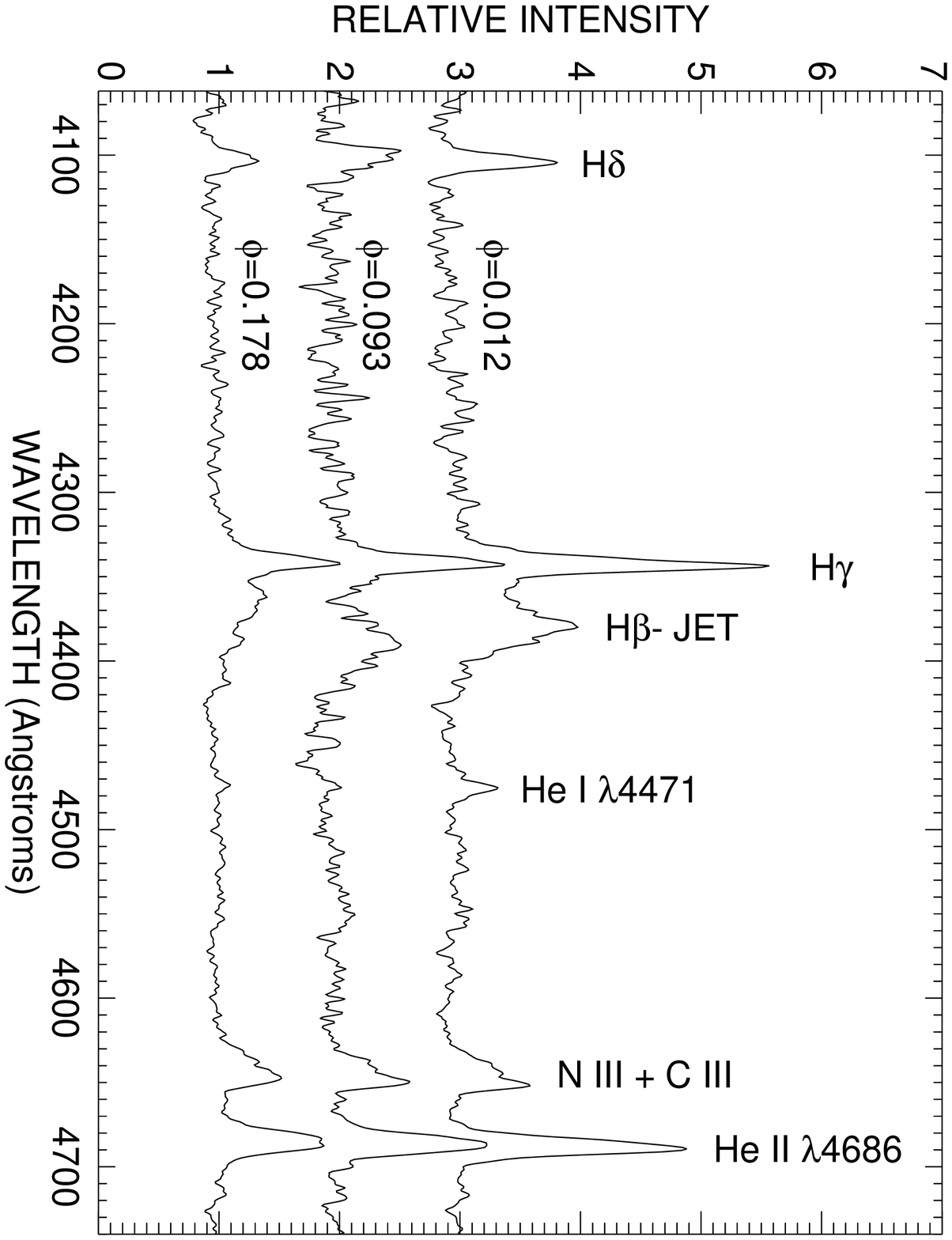}
\caption{}
\end{figure}
\clearpage

\begin{figure}[t]
\plotone{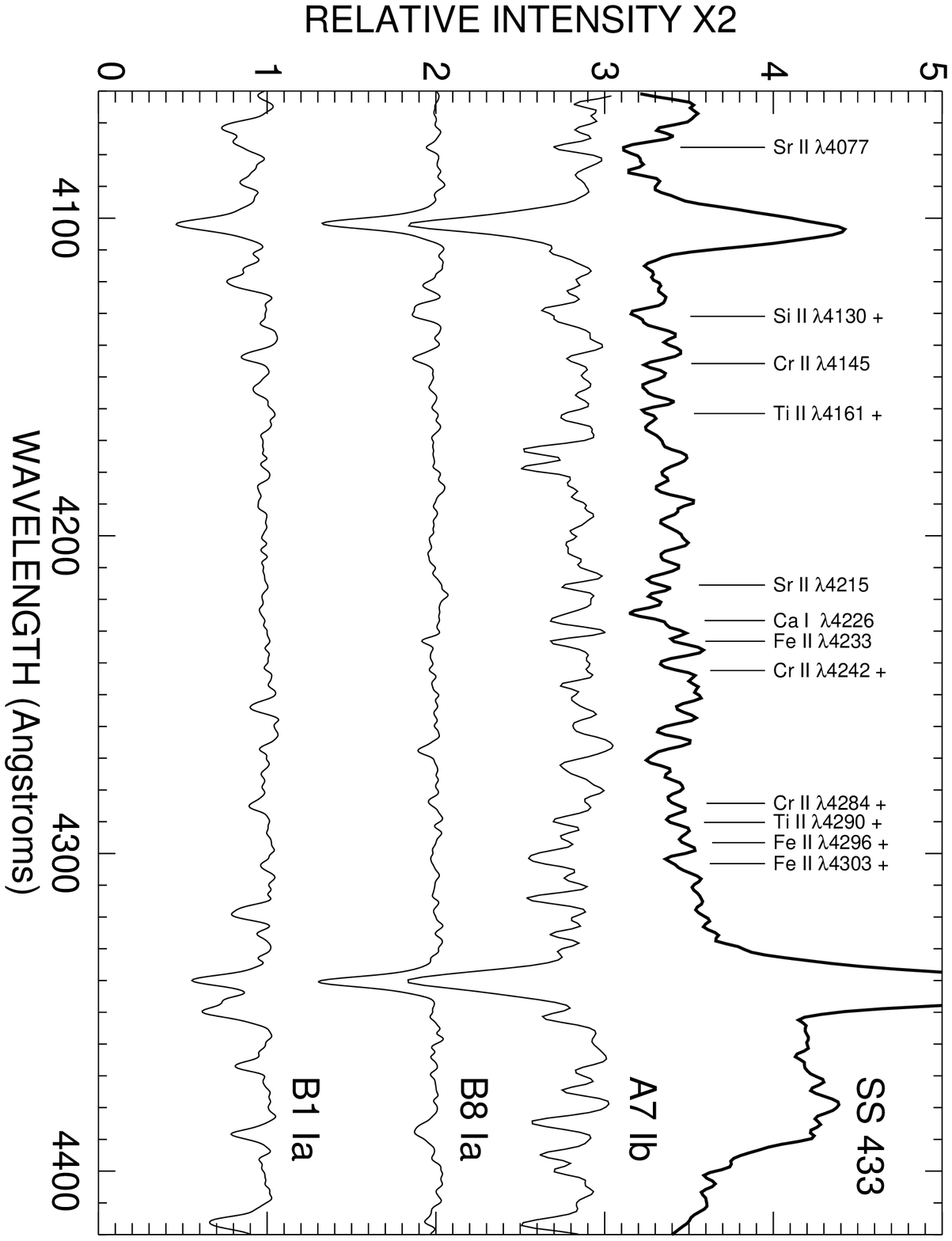}
\caption{}
\end{figure}
\clearpage

\begin{figure}[t]
\plotone{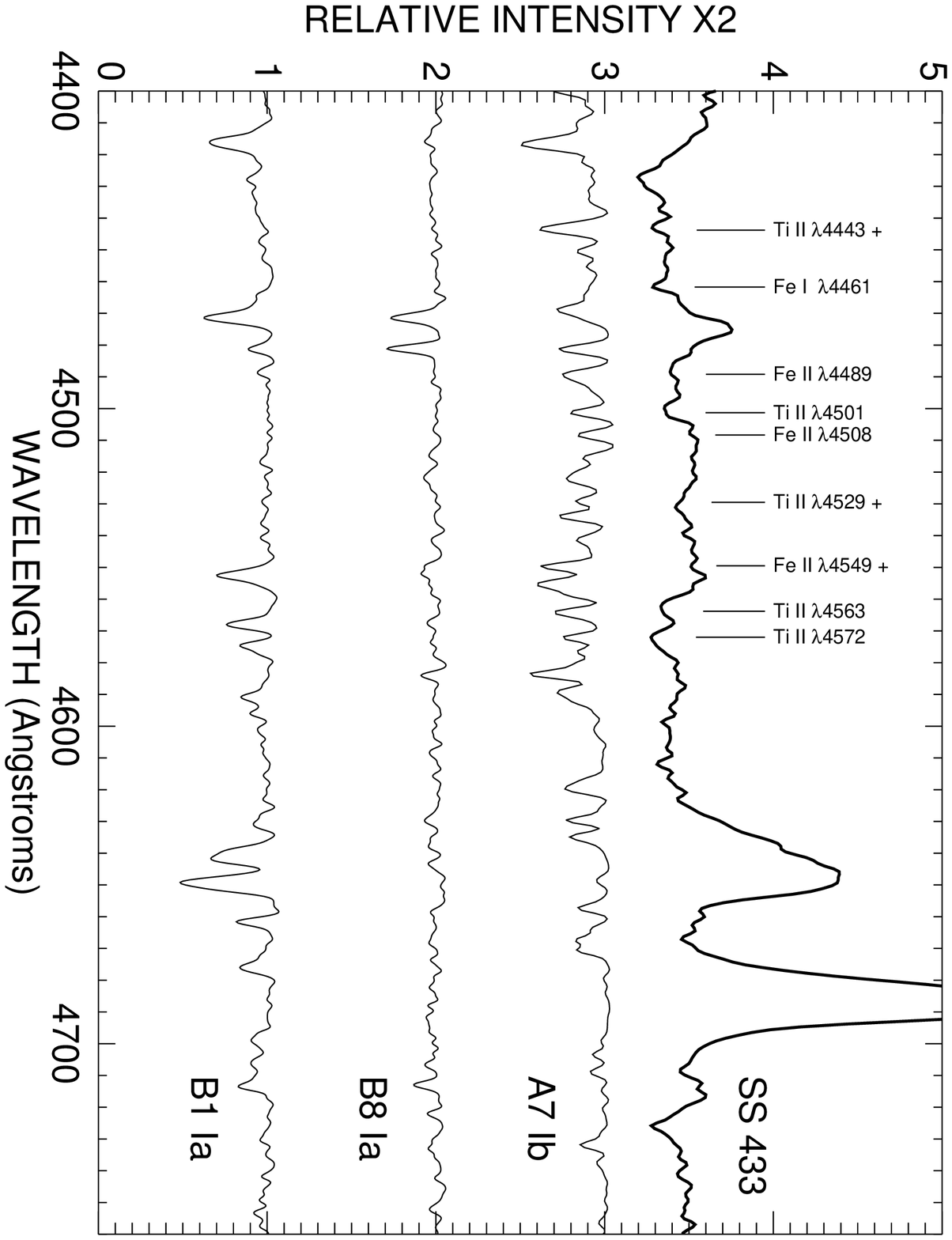}
\caption{}
\end{figure}


\end{document}